\documentstyle[12pt]{article}
\begin{document}
\begin{flushright}
{\normalsize
September 1996}
\end{flushright}
\vskip 0.2in
\begin{center}
\Large {Dissipative Effects in Photon Diagnostics of \\
 Quark Gluon Plasma }\\
\begin{center}
\normalsize {Sourav Sarkar$^a$, Pradip Roy$^a$, Jane Alam$^a$
, Sibaji Raha$^c$, and Bikash Sinha$^{a,b}$}\\
\end{center}
\vskip 0.2in
\small\it {a) Variable Energy Cyclotron Centre\\
     1/AF Bidhan Nagar, Calcutta 700 064\\
     India}
\vskip 0.15in
\small\it {b) Saha Institute of Nuclear Physics\\
           1/AF Bidhan Nagar, Calcutta 700 064\\
           India}
\vskip 0.15in
\small\it {c) Bose Institute\\
           93/1 A. P. C. Road, Calcutta 700 009\\
           India}

\end{center}

\addtolength{\baselineskip}{0.1\baselineskip} 
\parindent=20pt

\vskip 0.1in
\begin{center}
\bf {Abstract}\\
\end{center}
The effects of dissipation on the space time evolution of matter
formed in ultra-relativistic heavy ion collision is discussed.
The thermal photon spectra for RHIC and LHC energies with viscous
flow is considered.
The effects of viscosity in the thermal single photon spectra
is seen to be important in QGP
phase as compared to the hadronic phase. Recently available data from WA80
group at SPS energies for $S+Au$ collision are compared with theoretical
calculations. The experimental data do not appear to be 
compatible with
the formation of matter in the pure hadronic phase.
\newpage
\section*{I. Introduction}
   The primary motivation for studying heavy ion collisions at
ultrarelativistic energies is that such reactions provide an unique
opportunity to probe hadronic matter at very high temperature and/or
density. Numerical simulations of the QCD equation of state on the
lattice predict that at a sufficiently high temperature ($\sim 160
-200$) MeV, a phase transition should occur from the colour  confined
chirally asymmetric phase of QCD, the hadronic phase, to a (locally) colour
deconfined, chirally symmetric phase, referred to as the Quark-Gluon
Plasma(QGP) \cite {puri}. The QCD phase transition is of great relevance in
cosmology also \cite {bono}; within the standard big bang model, the
early universe should have experienced such a transition a few
microseconds after the big bang. Although the order of this phase
transition remains to be settled, a substantial volume of work has
already been carried out, assuming it to be of first order both in the
laboratory as well as in the early universe. In this work, we shall
tacitly assume that the hadron to QGP phase transition is indeed of
first order.

  In most of the works aimed at devising signals of the above mentioned
phase transition in the laboratory, the usual scenario is as follows:
after a canonical (proper)time $\sim 1$ fm/c (the strong
interaction time scale),
the excited matter within the reaction volume of the colliding heavy ions
comes to a (local) thermal equilibrium with a temperature $T_i$. If
this temperature $T_i$ is larger than the critical temperature $T_c$
of the QCD phase transition, then the excited matter would exist in the form of QGP. The QGP then evolves in space and time, cooling down to the
critical temperature $T_c$ when the confinement phase transition starts;
for a first order transition the system remains in a mixture of QGP
and hadronic matter at a temperature $T_c$ for some time the duration of which
is determined by the relative number of the degrees of freedom in the
QGP and hadronic phases. Such a configuration continues until all the
latent heat which maintains the temperature at $T_c$ by compensating
the cooling due to expansion is fully exhausted. At this stage all of
the QGP converts to hadronic phase and the system continues to cool
further, its dynamics being governed by the hadronic equation of state,
until the mean free paths of the constituent hadrons become too long to
maintain a collective behaviour. At this point, usually referred to as
freeze-out era, the momentum distributions of the particles become
frozen at the characteristic values when the particles suffer the last
collision, and the particles free stream thereafter, carrying the
space time integrated
informations to the detectors. For the sake of simplicity, most
studies on QGP diagnostics assume that the system behaves as an ideal
{\it i.e.} nondissipative, fluid throughout the entire evolution.

    Given the tremendous importance of the issue, it is imperative to
understand how far, if at all, such an idealisation is justified.
Even though a rigorous treatment of the dissipative effects in
relativistic fluid dynamics is still beset with several technical
as well as conceptual difficulties \cite{israel,strott}, one must admit that
in any realistic scenario, 
dissipative effects, in principle should have an important role.
In this work we consider the relativistic fluid to be weakly dissipative,
as has been argued to be the case by earlier authors\cite{danei}
and explore the ensuing consequences on the estimation of the initial
temperature and the photon spectra, data for which have become available of
late. Since our motivation in the present work  is to study the 
{\it relative} importance of the dissipative effect vis-a-vis an ideal
fluid dynamic scenario that has been routinely used by a number of
authors\cite{ja,kkmm,dksprc,dks} to compare the currently available data and/or
to estimate experimentally measurable quantities at RHIC or LHC, we adopt the 
same evolution scenario (Bjorken scaling flow), together with the same
formation time $\tau_i$ ($\approx$ 1 fm/c). It must be mentioned at
the outset that our present purpose is demonstrative rather than 
making firm phenomenological comparisons.

    The organisation of the present work is as follows. In the next section,
we briefly recapitulate the space-time evolution \cite{danei} of the system
with dissipation. In section III we discuss the effect of dissipation
on the boundary or initial conditions governing the evolution equation.
Section IV briefly recapitulates the formalism for photon diagnostics
of QGP. Finally in section V, we present the results of our actual
estimates at LHC, RHIC and SPS energies and compare them with the available
data\cite{wa80} at SPS. Section VI contains a brief summary and conclusions.

\section*{II. Viscous Hydrodynamics at Relativistic Energies}

   For a perfect fluid, the mean free path of the constituent particles
is negligibly small compared to all the available length scales
in the system. In a finite system like colliding nuclei, however large,
such a criterion may not be fulfilled and as such, the changes in the
collective quantities like pressure, energy density, number density, velocity
etc. over a distance of a mean free path cannot always be ignored.
This should lead to an inherent dissipative behaviour.

    In the presence of dissipation, the kinetic energy of the fluid decays as
heat energy. This
requires a redefinition of the energy-momentum tensor 
and the particle current
\cite{wein}. For the sake of brevity, we simply quote the result in the form
of energy-momentum conservation in an imperfect fluid which reads \cite{danei}
\begin{equation}
\frac{d\epsilon}{d\tau}\,+\,\frac{(\epsilon + P)}{\tau}=(\frac{4}{3}
\eta +\zeta)/{\tau^2}
\end{equation}
Eq.(1) is the scaling Navier-Stokes(NS) equation where we have assumed the
validity of the scaling solution \cite{matsui}, usually referred to as
the Bjorken solution \cite{bj}. The rate of heat generation
is $(4\eta/3+\zeta)/\tau^2$, which vanishes in the case of an ideal fluid
$(\eta\rightarrow 0, \zeta\rightarrow 0)$. Obviously, the occurrence
of the viscosity co-efficients $\eta$ and $\zeta$ in eq.(1) violates
the invariance under $\tau \rightarrow -\tau$, and consequently the
reversibility \cite{prigo}. As a result, entropy is generated during the
temporal evolution of the system; thus, estimating the initial
temperature from the final multiplicity in the usual manner \cite{ja}
constitutes an overestimate (see Appendix-I).

We assume, without any loss of generality
 that the functional relation between the
pressure and energy density in case of dissipative flow is
same as that of ideal flow \cite{wein}.
We shall, as is the usual practice, assume that the baryon
density is zero in the central region so that the thermal conductivity
$K=0$ \cite{danei} throughout.

\section*{IIa. Quark-Gluon Plasma Phase}

    The limits on the shear viscosity for the NS equation (eq. 1) in
the QGP phase have been discussed in depth in \cite{danei}(see also
\cite {hoso}).
The acceptable range of $\eta$ is
\begin{equation}
 2T^3 \leq \eta \leq 3T^3(\tau T)
\end{equation}
Obviously, the range is quite small unless the product $T\tau$ (very late
times and/or large temperatures) is very large. As should be clear
from what follows, such a situation is rather unlikely within the
QGP phase.

   The bulk viscosity $\zeta$, in general can be as large as $\eta$,
or in certain circumstances, even larger \cite {wein2}.
It was however shown by Weinberg \cite {wein2} some time ago
that if the trace of the
energy momentum tensor be expressible as a function of
$\epsilon$ and the pressure is the same function of $\epsilon$ as in
the adiabatic case, then $\zeta=0$. We thus obtain for the scaling
NS equation in the QGP phase
\begin{equation}
\frac{d\epsilon}{d\tau}+\frac{\epsilon+P}{\tau}=\frac{4}{3}\eta/\tau^2
\end{equation}
Although eq. (3) is somewhat simpler than eq. (1), we work with the
more general form for the time being. For $\epsilon=3a_QT^4$ and
$P=\epsilon/3$, where $a_Q=(\pi^2/90)g_{QGP}$, we can solve eq. (1)
with the boundary condition $T=T_i$ at $\tau=\tau_i$:
\begin{equation}
T=T_i(\tau_i/\tau)^{1/3}+(\frac{1}{6} \eta_{Q0}+\frac{1}{8} \zeta_{Q0})
\frac{1}{a_Q}\frac{1}{\tau_i}
\left[(\frac{\tau_i}{\tau})^{1/3}-\frac{\tau_i}{\tau}\right]
\end{equation}
where $\eta_{Q0}=\eta_Q/T^3$ and $\zeta_{Q0}=\zeta_Q/T^3$. The proper time
$\tau_Q$ taken by the system to cool down from the initial temperature
$T_i(>T_c)$ to the critical temperature $T_c$ can be calculated by
inverting eq. (4):
\begin{equation}
\tau_Q=\tau_Q(\tau_i, T_i, T_c)
\end{equation}
The life time of the QGP phase is $\tau_Q-\tau_i$. It can be readily
seen that in the presence of dissipation, the life time of the QGP
phase becomes longer (due to viscous heating)
relative to the ideal case, for the same values
of $T_i$, $\tau_i$ and $T_c$. For an actual estimate we use $\eta_{Q0}=2.5$
, as dictated by eq. (2) and $\zeta_{Q0}=0$ in the pure QGP phase.

\section*{IIb. The Mixed Phase}

    When the temperature of the system comes down to $T_c$, the system
goes over to a mixed phase composed of QGP and hadronic matter. In
this phase the temperature remains constant but the energy density
changes due to the change in the statistical degeneracy on account
of hadronization. We define $f_Q$ as the volume fraction of the
QGP in the mixed phase, so that
the energy density at any proper time
$\tau$ during the life time of the mixed phase is given by
\begin{equation}
\epsilon(\tau)=f_Q(\tau)\epsilon_Q+(1-f_Q(\tau))\epsilon_H
\end{equation}
where $\epsilon_Q(\epsilon_H)$ is the energy density in the
QGP(hadronic) phase at temperature $T_c$.
Eq.(6) is obtained by a linear interpolation between the two
points ($1,\epsilon_Q$) and ($0,\epsilon_H$) in
the ($f_Q,\epsilon $) plane.
For the coefficients of viscosity, we can formally use a similar
interpolation,
\begin{equation}
\eta(\tau)=f_Q(\tau)\eta_Q+(1-f_Q(\tau))\eta_H
\end{equation}
and
\begin{equation}
\zeta(\tau)=f_Q(\tau)\zeta_Q+(1-f_Q(\tau))\zeta_H
\end{equation}
where the subscript $Q(H)$ denotes the value in the QGP(Hadronic) phase
at $T_c$. The applicable limits on $\eta_{Q0}$ and $\zeta_{Q0}$
have already been discussed in the previous section. The corresponding
situation in the hadronic phase is explained in the following section.

  The evolution equation for $f_Q(\tau)$ in the mixed phase
can be readily obtained from eqs. (1), (6), (7) and (8):
\begin{equation}
\frac{df_Q}{d\tau}+(\frac{1}{\tau}-\frac{b}{\tau^2})f_Q=
\frac{c}{\tau^2}-\frac{a}{\tau}
\end{equation}
where $a=(\epsilon_H+P_c)/{\Delta\epsilon}$ ;
$b=4(\eta_Q-\eta_H)/{3\Delta\epsilon}$ ;
$c=(2\xi_0+\frac{4}{3}\eta_H)/{\Delta\epsilon}$ ;
$\Delta\epsilon=\epsilon_Q-\epsilon_H$.
The solution of this equation with the boundary condition $f_Q=1$ at
$\tau=\tau_Q$ is given by
\begin{equation}
f_Q=(ab-c)\left[Ei(b/\tau)-Ei(b/\tau_Q)\right]
\frac{e^{-b/\tau}}{\tau}
+(a+1)\frac{\tau_Q}{\tau}e^{(b/\tau_Q-b/\tau)}-a
\end{equation}
where $Ei(x)$ is the exponential function. For $b=c=0 (\eta=\xi=0)$
we recover the solution for $f_Q$ \cite{ja} in case of a perfect fluid.
Note however that the bulk viscosity $\zeta$ in the mixed phase may be
non-zero due to the change in sound velocity over a finite relaxation
time \cite{land}. We use $\zeta=2\eta/3$, following Hosoya and Kajantie
\cite{hoso}. The time $\tau_H$, when the mixed phase ends, is obtained from
eq. (10) by demanding $f_Q(\tau_H)=0$. We thus have,
\begin{equation}
\tau_H=\tau_H(\tau_Q)
\end{equation}
for given equations of state in the QGP and the hadronic phases.
The life time of the mixed phase
is then $\tau_H-\tau_Q$.

\section*{IIc. Hadronic Phase}

   The pure hadronic phase starts at $T=T_c$ at the time $\tau_H$ so that
the solution of eq. (1) in the hadronic phase is again eq. (4) with
the appropriate substitution of the $a_Q$ by $a_H$, $T_i$ by $T_c$
and $\tau_i$ by $\tau_H$:
\begin{equation}
T=T_c(\tau_H/\tau)^{1/3}+(\frac{1}{6} \eta_{H0}+\frac{1}{8} \zeta_{H0})
\frac{1}{a_H}\frac{1}{\tau_H}
\left[(\frac{\tau_H}{\tau})^{1/3}-\frac{\tau_H}{\tau}\right]
\end{equation}
In the hadronic phase, the co-efficient of the shear viscosity
can be obtained from the estimated transport cross-section
according to the relation $\eta_H\approx T/\sigma_{tr}\sim
(T/200 MeV)\,(0.5-1)/$fm$^3$\cite{danei}. The expected range of
$\sigma_{tr} \sim 10-20$ mb  which corresponds to $\eta_{H0} \sim
0.75-1.5$ at $T_c$.
In what follows we shall use the larger value of 1.5
for $\eta_{H0}$ as a conservative case.
 The value of the bulk
viscosity can be non-zero in the hadronic phase due to lack
of chemical equlibrium and its consequence on the sound velocity
\cite{degroot}, we again take $\zeta_H=2\eta_H/3$ \cite{hoso}.

    As mentioned in Section I, the system freezes out at a time denoted by
$\tau_f$, when it becomes too dilute to maintain a collective flow.
There is as yet no universal criterion for determining the instant
of freeze-out, for a detailed discussion, see \cite{ja}. In the
present context, we describe freeze-out through a freeze-out temperature
$T_f$ which we treat as a free parameter  ($\le 140$ MeV). Then
from eq.(12) $\tau_f$
is given by
\begin{equation}
\tau_f=\tau_f(\tau_H, T_c, T_f)
\end{equation}
so that the life time of the hadronic phase is $\tau_f-\tau_H$.

\section*{III. Hydrodynamic Flow and Connections with Observables}

   In the case of an ideal fluid, the conservation of entropy
implies that the rapidity density $dN/dy$ is a constant of motion
for the isentropic flow \cite{bj}.
In such circumstances, the experimentally
observed multiplicity, $dN/dy$ may be related to
a combination of the initial
temperature $T_i$ and the initial time $\tau_i$ as $T_i^3\tau_i$.
Assuming an appropriate value of $\tau_i$(taken to be $\sim 1$ fm/c),
one can estimate $T_i$.

   For dissipative systems, such an estimate is obviously inapplicable
(see appendix I).
Generation of entropy during the evolution invalidates the role of
$dN/dy$ as a handy constant of motion. Moreover, the irreversibility
arising out of dissipative effects implies that estimation~ of the
~initial ~temperature from the final~ rapidity density~ is no longer
a trivial task. We can, nevertheless, relate the ~experimental
$dN/dy$ to the freeze-out temperature $T_f$ and the freeze-out
time $\tau_f$ by the relation,
\begin{equation}
\frac{dN}{dy}=\pi R_A^2 4 a_H T_f^3 \tau_f/c
\end{equation}
where $R_A$ is the radius of the colliding nuclei(we consider $A A$
collision for simplicity) and c is a constant $\sim 3.6$.
The parameter $a_H$ occurring in
eq. (14) is the so called statistical degeneracy ($a=(\pi^2/90)g$)
for an effectively massless hadronic gas. In this work we assume
the hadronic matter to consist of $\pi$, $\rho$, $\omega$ and $\eta$,
which allows us to estimate $a_H$ through a parametrization introduced
earlier \cite{some}. It must also be mentioned that eq. (14) still
constitutes a tremendous amount of idealization; for a discussion of
all the caveats, see\cite{ja}. For our present purpose, we assume that
eq. (14) does hold at the freeze-out era. 
We shall also neglect
transverse expansion in this work,
which would amount to an 
overestimation of $\tau_f$, for a given $T_f$ and
$dN/dy$. We however do not consider this to be 
a major factor; this is to be discussed in some detail in Sec. V.

\section*{IV. Photon Spectra}

In this section we will briefly discuss the photon spectra
due to an expanding quark gluon plasma.
In the QGP phase the main contribution comes from the
annihilation ($q\bar q \rightarrow g\gamma$) and Compton processes
($q(\bar q)g \rightarrow q(\bar q)\gamma$). In the hadronic phase
(composed of  $\pi$, $\rho$, $\omega$, $\eta$), an array of reactions like
$\pi\rho \rightarrow
\pi\gamma$, $\pi\eta \rightarrow \pi\gamma$,
$\pi\pi \rightarrow \eta\gamma$, $\pi\pi \rightarrow \rho\gamma$,
$\pi\pi \rightarrow \gamma\gamma$
and the decays $\rho^0 \rightarrow \pi^+\pi^-\gamma$ and $\omega \rightarrow
\pi^0\gamma$ are considered \cite{kapusta}.

The rate of emission of photons per unit volume from QGP phase is given
by \cite{kapusta},
\begin{equation}
E\frac{dR}{d^3p}=\frac{5}{9}\,\frac{\alpha\alpha_s}{2\pi^2}T^2
e^{-E/T}\,ln\left(\frac{2.912}{g^2}\,\frac{E}{T}+1 \right)
\end{equation}
For our exploratory calculations we will consider the rates of photon
emission from the QGP and the hadronic matter to be equal as in \cite{kapusta,
ja2}. It is a very useful first approximation
for identifying the $p_T$ window where we concentrate on the
signature of QGP. The $p_T$ distribution of photons is obtained
by convoluting the basic rates with the space time history,
\begin{eqnarray}
\frac{dN}{d^{2}p_{T}dy}&=&\pi R_A^2\int\left[\left(E\frac
{dR}{d^{3}p}\right)_{QGP} \Theta(\epsilon-\epsilon_{Q})\right.\nonumber\\
& &+\left\{\left(E\frac{dR}{d^{3}p}\right)_{QGP}f_Q
+\left(E\frac{dR}{d^{3}p}\right)_{H}(1-f_Q)\right\}
\Theta(M)\nonumber\\
& &+\left.\left(E\frac{dR}{d^{3}p}\right)_{H}
\Theta(\epsilon_{H}-\epsilon)\right]
\tau\,d\tau \,d\eta
\end{eqnarray}
where $\Theta(M)=\Theta(\epsilon_Q-\epsilon)
\Theta(\epsilon-\epsilon_H)$.
The above equation along with eqs. (4), (10) and (12) is used to
evaluate the photon spectra.

\section*{V. Results}

    As we  have mentioned in section III, ~estimation of
the ~initial energy density~/temperature from the final state ~rapidity
density is no longer a trivial task in the presence of ~dissipation. The
principle however is quite straight forward. In our algorithm, we tacitly
assume that the initial formation time $\tau_i$ of the thermalised system
, be it in the QGP phase or the hadronic phase, has the canonical value
of $1$ fm/c. That the initial formation time (and consequently the initial
temperature $T_i$) is a poorly known quantity is beyond dispute. 
Indeed, it is now widely believed \cite{prl,Shur,CRS} that at RHIC
or LHC, the gluons may equilibrate considerably earlier than 1 fm/c.
However, even at these energies, the u/d quarks appear to 
equilibrate only $\sim$ 1 fm/c \cite{prl}.
Moreover,
all the works aimed at comparing the experimental data with theoretical 
predictions take, $\tau_i$ to be, as a rule, equal to or greater than 1 fm
\cite{ja,dksprc,dks,bj,Fai}.
Most of these works also use the Bjorken scaling hydrodynamics to
emulate the space-time evolution of the system. For a meaningful 
comparison with these works,
we should also therefore adopt the same scenario \footnote{ The
assumption of 1 fm/c as the initial formation time need not be taken
seriously. For our present purpose, one can always adopt the
viewpoint that we compare the temperature at a proper time 1 fm/c, a
sufficiently early instant of time in the evolution history, in the
dissipative vis-a-vis ideal fluid scenarios.}.

   We treat $T_i$ as a parameter; for each $T_i$, we let the system evolve
forward in time under the condition of dissipative fluid dynamics (eq. (1))
till a given freeze-out temperature $T_f$ is reached.
Thus $\tau_f$ is determined. We then compute $dN/dy$ at this instant of time
from eq. (14) and compare it with the experimental $dN/dy$. The
value of $T_i$ for which the calculated $dN/dy$ matches the
experimental number is taken to be the approximate initial temperature.
Schematically the calculation for $T_i$ proceeds (forward in time) as follows.
\begin{equation}
\tau_Q(T_i)\rightarrow \tau_H(\tau_Q(T_i))\rightarrow \tau_f(\tau_H(\tau_Q
(T_i)))
\end{equation}
i.e. $\tau_H$ depends on $T_i$ through $\tau_Q$ and finally $\tau_f$
depends on $T_i$ through $\tau_H$ ($cf$ eqs. (5),(11) and (13)). We then
solve the non-linear eq. (14) to estimate $T_i$ for given values of
$dN/dy$ and $T_f$.
We adopt a truely conservative view point in the sense that we
start with values of $T_i$ less than $T_c$ so that the system
is formed in the hadronic phase, and allow values of $T_i\geq T_c$
only if the discrepancy between the calculated and experimental
(or estimated) values of $dN/dy$ necessarily demands it. In these
circumtances, we naturaly let the system evolve first in the quark
gluon plasma phase, then the mixed phase and the hadronic phase.

    The result of these calculations are summarized in table I.
 For the sake of comparison, we also list the corresponding values
 as calculated in the ideal case (superscript Bj). For RHIC and LHC
 , the experimental $dN/dy$ values are the usual estimates obtained
 from extrapolating the pp data\cite{satz}. For SPS
 , the $dN/dy$ is known from actual measurement\cite{albrecht}.

    It can be readily seen from table I that in all cases, the initial
temperatures for the dissipative case are
lower than in the ideal case.
Lifetimes for the dissipative case
are also comparatively longer. One also finds that the lower the
freeze-out temperature, the lower the initial temperature although
the dependence is rather weak.

    It is evident from Table I that both in RHIC and LHC, the initial
    temperature is substantially higher than $T_c$ even for
    dissipative dynamics. (We use the canonical value of $160$ MeV
for $T_c$ in this work.)

    The situation at SPS demands special discussion. Please therefore
    refer to Table 2a. We find that an initial QGP phase may be formed
only if we use $T_c$ lower than $160$ MeV. For example, $T_c$ of $150$
 MeV  allows a QGP phase with an initial temperature of $154$ MeV  only,
and that too for a freeze-out temperature of $100$ MeV. Thus an
unambiguous conclusion about the formation of the QGP on the basis
of presently available SPS data appears to be doubtful. We therefore
explore the situation in somewhat greater detail in Table 2b.

   In Table 2b we study, on the one hand, the possibility that
the system is formed entirely in the hadronic phase with dissipative
dynamics. We look at the most conservative scenario of very low
freeze-out temperature, consistent with the argument of Hama
and Navarra\cite{hama} for the case of $S+Au$ system.
We find that to obtain a $dN/dy$ of $225$, an initial temperature
of $262$ MeV  is required which is obviously much larger than
the usual $T_c$ of $160$ MeV  (or even $200$ MeV ). It is thus
fair to conclude that for a reasonable formation time of $1$ fm/c
, the present SPS data are not compatible with a purely hadronic
phase even if dissipative effects are taken into account.
This, together with the results detailed in Table 2a makes the
situation most interesting indeed. We thus explore the remaining
possibility that the initial system is formed in the mixed phase,
{\it i. e.}, $T_i=T_c$.

     Table 2b shows that this case fits the data reasonably well. Depending
     on the choice of $T_f$, the initial system could be  a mixture
     of $80-90\%$ quark matter and the rest hadronic matter,
     co-existing in thermal equilibrium at a temperature $T_c$.
It thus appears that there is a strong hint of deconfinement
phase transition of first order in the currently available data
but for a clear indication of the formation of the QGP phase
, one must go to higher energies at RHIC or LHC. This situation
is also similar to that advocated by Werner\cite{werner} of late, where
he suggests the formation of quark blobs in hadronic matter at
SPS energies.

     Effects of these considerations on QGP diagnostics are of direct
     relevance in the present scenario. One of the promising
signals of QGP formation is direct photons \cite{ja,dks}.
We thus show in figs. 1 and 2 the calculated photon spectra for ideal
as well as dissipative fluid dynamics at RHIC and LHC energies
respectively. The immediate feature that stands out from these
figures is that the effect of dissipation is stronger in the QGP
phase than in the hadronic phase. This is to be expected as the
temperature in the QGP phase is higher, corresponding to the
larger values of the co-efficients of viscosity. It should also be
noted that even if the estimate of $T_i$ is lower than the ideal
case by only 10$\%$ or less, the change in the photon spectra at
high $p_T$ is quite sizeable.

   In fig. 3, we compare the photon spectra corresponding to the
configuration of Table 2b (initial mixed phase at a temperature
of $160$ MeV ) with the experimental data of WA80\cite{wa80,santo}.
For the sake of comparison,
we also show the photon spectra from purely hadronic phase with the
initial configuration of Table 2b. Obviously, the agreement with
the initial mixed phase is better, considering the fact that
the experimental data gives the upper limit for the photon spectra
(fig.3). Before closing this section, we feel that afew comments on
our neglect of transverse expansion are in order. As we already
mentioned, this amounts to an overestimate of $\tau_f$, especially
for RHIC and LHC energies. It is well known that the transverse flow
effects at SPS energies are not at all large \cite{ja2}.
Nevertheless, an inspection of Table 1. reveals that the change due
to dissipative effects in total lifetime or $T_i$ relative to the
ideal case is less than 10$\%$ at both RHIC and LHC energies. Since
the same effect is seen both in $\tau_f$ and $\tau_H$, it is
reasonable to expect that even in the presence of transverse flow,
the same qualitative features would obtain. This however ought to be
verified in a full (3+1) dimensional calculation which we plan to
perform in the near future.

\section*{VI. Summary}

 In this work, we have explored the effects of dissipation in the
space time evolution of matter formed in very energetic collisions
of heavy ions. We find that even with the assumption of scaling
hydrodynamic flow and weak dissipation, the deviation from the ideal
case is not negligible. The presence of dissipation reduces the rate
of cooling, resulting in a longer total lifetime. For a given total
multiplicity in the final state, the estimated initial temperature
for the dissipative case is smaller than in the ideal situation.
This effect is most pronounced at SPS energies. It appears that the
presence of dissipation at RHIC or LHC energies may not change the
estimated $T_i$ to such an extent as to qualitatively invalidate the
conclusions derived from the assumption of ideal hydrodynamics. At
the level of the present analysis, we find that the measured $dN/dy$
of 225 cannot be easily reproduced without a first order phase
transition (Table 2b). The most likely situation seems to be an
initial configuration of a mixed state of hadronic and quark matter.

  In the tharmal single photon spectra, we observe that the effect
of dissipation should show up at the high $p_T$ sector. This is
quite natural, given the fact that dissipation results in a lower
initial temperature. In all cases, the effect of viscosity is
important in the QGP phase as compared to the hadronic phase.
Recently
available data from the WA80 group at SPS energies for S + Au
collisions have been compared with theoretical calculations without
any additional free parameters. we find that the parameters
extracted from the $dN/dy$ analysis (Table 2b) yield a reasonably
good fit
to the data, if the initial configuration is indeed a mixed phase.
The data are not compatible with formation of matter in a pure
hadronic phase. These conclusions are not expected to be materially
alterred due to transverse flow which we have 
not taken into account for the
present.

\newpage

\section*{Table Captions}
\begin{itemize}
\item{Table 1:} Comparison of time scales  for the different phases
and initial temperatures between Bjorken hydrodynamics(superscript Bj)
and viscous hydrodynamics at RHIC and LHC energies.
$dN/dy$ is the multiplicity,
$T_i$, $T_c$ and $T_f$ are the initial, critical and freeze-out
temperatures respectively. $\tau_i$,$\tau_Q$, $\tau_H$ and $\tau_f$
are the initial thermalisation time, time when phase transition starts,
time when mixed phase ends and freeze-out time rescpectively.

\item{Table 2a:} Same as Table 1 at SPS energies. The value of $dN/dy$
is taken from experiment.

\item{Table 2b:} Same as Table 2a when matter is formed in pure hadronic or
mixed phase.
\end{itemize}

\section*{Figure Captions}

\begin{itemize}

\item{Figure 1:} Photon~ spectrum~ at RHIC energies~ for ideal and~
~viscous flow.  The inputs are taken~ from table.1, with freeze-out~
~temperature $T_f=100$ MeV. QM(HM)~ denotes the~ contributions from~
~pure QGP(hadronic) and~ QGP(hadronic) part of the~ mixed phase~.

\item{Figure 2:} Photon spectra at LHC energies with
and without viscous flow.

\item{Figure 3:} Photon Spectra at SPS energies with viscous flow and
comparison with WA80 data. The dashed curve denotes the photon
spectrum with inputs from no phase transition scenarion of table 2b.
The solid curve denotes the photon spectrum with inputs from table
2b. The freeze-put temperature is taken as $65$ MeV as calculated
by applying the method of ref.\cite{hama}. The arrow denotes the experimental
data from WA80\cite{santo,wa80}.
\end{itemize}
\newpage
\section*{APPENDIX-I}

In case of ideal fluid, the conservation of entropy implies
that the rapidity density $dN/dy$ is a constant of motion.
In such cases the experimentally measured value of the final rapidity
density, which is a quantity depends on the integration over
the space time history, can be connected  to the initial temperature.
In these calculations one assumes the validity of the
extrapolation from the
freeze point (with temperature $T_f$ and time $\tau_f$)
to the initial point specfied by
temperature $T_i$ and proper time $\tau_i$ backward in time.
But in the presence of dissipative effects the process is irreversible,
and one shoud take care of this fact.
In this appendix we will show that if we solve the eq.(1) backward in time
then we always overestimate the initial temperature for a given
vaule of $dN/dy$ and $\tau_i$.

Under the transformation $\tau\rightarrow -\tau$,
eq.(1) becomes,
$$\frac{d\epsilon}{d\tau}\,+\,\frac{(\epsilon + P)}{\tau}=-(\frac{4}{3}
\eta +\zeta)/{\tau^2} \eqno {(A.1)}$$	
For the equation of state  of a massless gas,
eq.(15) can be written as
$$\frac{d}{d\tau}(T\tau^{1/3})=-\frac{b}{12a_k}\frac{1}{\tau^{5/3}} \eqno
{(A.2)}$$
where $\eta=\eta_0T^3$, $\zeta=\zeta_0T^3$,
$b=\frac{4}{3}\eta_0+\zeta_0$, $a_k=(\pi^2/90)g_k$, $g_k$
is the statistical degeneracy.
The solution of eq.(A.2)  with the boundary condition $T=T_f$
at $\tau=\tau_f$ can be written as,
$$T\tau^{1/3}=
K\left(dN/dy\right)^{1/3}-\frac{b}{8a_k}\left(\tau_f^{-2/3}
-\tau^{-2/3}\right) \eqno {(A.3)}$$
where we have  used the relation $T_f\tau_f^{1/3}=K(dN/dy)^{1/3}$.
The initial temperature estimated from eq.(A.3) is
$$T_i^{\prime}\tau_i^{1/3}=
K\left(dN/dy\right)^{1/3}-\frac{b}{8a_k}\left(\tau_f^{-2/3}
-\tau_i^{-2/3}\right) \eqno {(A.4)}$$
$T_i^{\prime}$ denotes the initial temperature when we solve the
hydrodynamic equation backward in time.
Similarly we can solve the
hydrodynamic eq.(1) with boundary condition $T=T_i$ at
$\tau=\tau_i$, the solution is given by,
$$T\tau^{1/3}=
T_i\tau_i^{1/3}
+\frac{b}{8a_k}\left(\tau_i^{-2/3}-\tau^{-2/3}\right) \eqno {(A.5)}$$
At the freeze-out time eq.(A.5) can be written in terms of $dN/dy$,
$$T_i\tau_i^{1/3}=
K\left(dN/dy\right)^{1/3}+\frac{b}{8a_k}\left(\tau_f^{-2/3}
-\tau_i^{-2/3}\right) \eqno {(A.6)}$$
In case of ideal hydrodynamics,($b=0$)
$T_i^{\prime}=T_i$.
But in case of viscous flow for a given $dN/dy$ and $\tau_i$
$$T_i^{\prime}>T_i \eqno {(A.7)}$$
\newpage

\begin{center}
{\bf {\underline {Table 1}}}
\vskip 0.3 in
\begin{tabular}{|cccccccc|}
\hline
 & & & & & & &\\
    & $T_f$ & $T_c$ & $\tau_i/\tau_i^{Bj}$ & $\tau_Q/\tau_Q^{Bj}$ & $\tau_H/\tau_H^{Bj}$ & $\tau_f/\tau_f^{Bj}$ & $T_i/T_i^{Bj}$  \\
 & (MeV) & (MeV) & (fm/c) & (fm/c) & (fm/c) & (fm/c) & (MeV) \\
 & & & & & & &\\
\hline
 & & & & & & &\\
RHIC & 100 & 160 & $1/1$ & $3.5/4.2$ & $35/34$ & $150/138.4$ & $225/258$\\
$(\frac{dN}{dy}=1735)$ & 140 & 160 & $1/1$ & $3.6/4.2$ & $36/34$ & $54.7/50.5$ & $228/258$ \\
 & & & & & & &\\
LHC & 100 & 160 & $1/1$ & $13.7/13.6$ & $117.2/109.8$ & $486.8/449.8$ & $358/382$\\
$(\frac{dN}{dy}=5624)$ & 140 & 160 & $1/1$ & $13.8/13.6$ & $118.2/109.8$ & $177.4/163.9$ & $359/382$ \\
 & & & & & & &\\
\hline
\end{tabular}
\end{center}

\newpage

\begin{center}
{\bf {\underline {Table 2a}}}
\vskip 0.5 true in.
\begin{tabular}{|cccccccc|}
\hline
 & & & & & & &\\
    & $T_f$ & $T_c$ & $\tau_i/\tau_i^{Bj}$ & $\tau_Q/\tau_Q^{Bj}$ & $\tau_H/\tau_H^{Bj}$ & $\tau_f/\tau_f^{Bj}$ & $T_i/T_i^{Bj}$  \\
 & (MeV) & (MeV) & (fm/c) & (fm/c) & (fm/c) & (fm/c) & (MeV) \\
 & & & & & & &\\
\hline
 & & & & & & &\\
SPS & 100 & 120 & $1/1$ & $3.5/1.7$ & $37.9/13.6$ & $67.5/23.5$ & $165/203$\\
$(\frac{dN}{dy}=225)$ & 100 & 150 & $1/1$ & $1.1/2.5$ & $18/20$ & $67.5/67.5$ & $154/203$ \\
 & & & & & & &\\
\hline
\end{tabular}
\vskip 1.0 true in
{\bf {\underline {Table 2b}}}
\vskip 0.5 in
\begin{tabular}{|ccccccc|}
\hline
 & & & & & & \\
No & Phase & Transition & & & &\\
 & & & & & & \\
\hline
 & & & & & & \\
$\frac{dN}{dy}$ & $T_i$ & $\tau_i$ & $T_f$ & $\tau_f$ & & \\
 & (MeV) & (fm/c) & (MeV) & (fm/c) & & \\
 & & & & & & \\
\hline
 & & & & & & \\
225 & 264 & 1 & 65 & 246 & &\\
 & & & & & & \\
\hline
 & & & & & & \\
 Formation & in & Mixed & Phase & & & \\
 & & & & & & \\
\hline
 & & & & & & \\
$\frac{dN}{dy}$ & $T_f$ & $T_c(=T_i)$ & $\tau_Q(=\tau_i)$ & $\tau_H$ & $\tau_f$ & $f_0$  \\
 & (MeV) & (MeV) & (fm/c) & (fm/c) & (fm/c) & \\
 & & & & & & \\
\hline
 & & & & & & \\
225 & 140 & 160 & 1 & 16 & 25 & 0.92 \\
225 & 100 & 160 & 1 & 15 & 68 & 0.81 \\
225 & 65 & 160 & 1 & 14 & 247 & 0.75 \\
 & & & & & & \\
\hline
\end{tabular}
\end {center}

\newpage
\begin{thebibliography}{39}


\bibitem{puri} Proc. of Winter School on Quark Gluon Plasma, Dec.1989
, Puri, India, eds. B. Sinha, S. Pal, and S. Raha, Springer-Verlag
Germany.

\bibitem{bono} S. A. Bonometto and O. Pantano, Phys. Rep. {\bf 228}, 175(1993).

\bibitem{israel} W. Israel and J. M. Stewart, Proc. of Roy. Soc. (London)
{\bf 365}, 43(1979).

\bibitem{strott} D. Strottman, Nucl. Phys. {A566}, 245c(1994).

\bibitem{danei} P. Danielewicz and M. Gyulassy, Phys. Rev. D {\bf 31},
53 (1985).

\bibitem{ja} J. Alam, S. Raha and B. Sinha, Phys. Rep. (in press)

\bibitem{kkmm} R. C. Hwa and K. Kajantie, Phys. Rev. {\bf D 32}, 1109(1985),
K. Kajantie, J. I. Kapusta, L. McLerran, and A. Mekjian, Phys. Rev.
{\bf D 34}, 2746(1986), D. K. Srivastava, B. Sinha, M. Gyulassy, and
X. -N. Wang, Phys. Lett. {\bf 276}, 285(1992). 

\bibitem{dksprc} D. K. Srivastava, B. Sinha, and C. Gale, Phys. Rev.
{\bf C 53}, R567(1996).

\bibitem{dks} D. K. Srivastava and B. Sinha, Phys. Rev. Lett.
{\bf 73}, 2421 (1994).

\bibitem{wa80} R. Albrecht et al, ``Limits on Production of Direct
Photons in $200\cdot$A GeV $^{32}$S+Au Collisions.'' to appear in Phys.
Rev. Lett.

\bibitem{wein} S. Weinberg, {\it Gravitation and Cosmology}, John
Willey and Sons, New York 1972.

\bibitem{matsui} M. Gyulassy and T. Matsui, Phys. Rev. D {\bf 11},
192 (1984).

\bibitem{bj} J. D. Bjorken, Phys. Rev. D {\bf 27}, 140 (1983).

\bibitem{prigo} I. Prigogine, {\it Introduction to Thermodynamics
of Irreversible Processes}, Interscience Publications (NY, 3rd
ed.).

\bibitem{hoso} A. Hosoya and K. Kajantie, Nucl. Phys. B {\bf
250}, 666 (1985).

\bibitem{wein2} S. Weinberg, Astrophysical Journal {\bf 168}, 175
(1971).

\bibitem{land} L. D. Landau and E. M. Lifshitz, Fluid Mechanics,
Pregamon Press, Oxford, 1959.

\bibitem{degroot} S. R. DeGroot, Acta physica Austrica Suppl. X, 529
(1973).

\bibitem{some} S. Chakraborty, J. Alam, D. K. Srivastava, B.
Sinha and S. Raha Phys. Rev. D{\bf 46} 3802 (1993).

\bibitem{kapusta} J. Kapusta, P. Lichard and D. Seibert,
Phys. Rev. D {\bf  44}, 2774(1991).

\bibitem{ja2} J. Alam, D. K. Srivastava, B. Sinha and D. N. Basu,
Phys. Rev. D {\bf  48},1117(1993).

\bibitem{prl} J. Alam, S. Raha, and B. Sinha, Phys. Rev. Lett. 
{\bf 73}, 1895(1994). 
 
\bibitem{Shur} E. Shuryak, Phys. Rev. Lett. {\bf 68}, 3270(1992).

\bibitem{CRS} S. Chakrabarty, S. Raha, and B. Sinha, Mod. Phys.
Lett. A {\bf 7}, 927(1992).

\bibitem{Fai} J. J. Neumann, D. Seibert, and G. Fai, Phys. Rev. {\bf 
C 51}, 1460(1995).

\bibitem{satz} H. Satz, in Proc. of ECFA Large Hadron Collider
Workshop, Aachen, Germany edited by G. Jarlskog and D. Rein
CERN rep. no. 90-10, Geneva, vol. I, p188 (1990).

\bibitem{albrecht} R. Albrecht et al, Z. Phys. {\bf C55}, 539(1992).

\bibitem{hama} Y. Hama and F. S. Navarra, Z. Phys. {\bf C 53}, 501(1992).

\bibitem{werner} K. Werner, Phys. Rev. Lett. {\bf 73},1594(1994).

\bibitem{santo} R. Santo et al. Nucl. Phys. A {\bf 566} 61c, (1994);
Report No. IKP-MS-93/0701, Muenster, 1993.


\end{thebibliography}
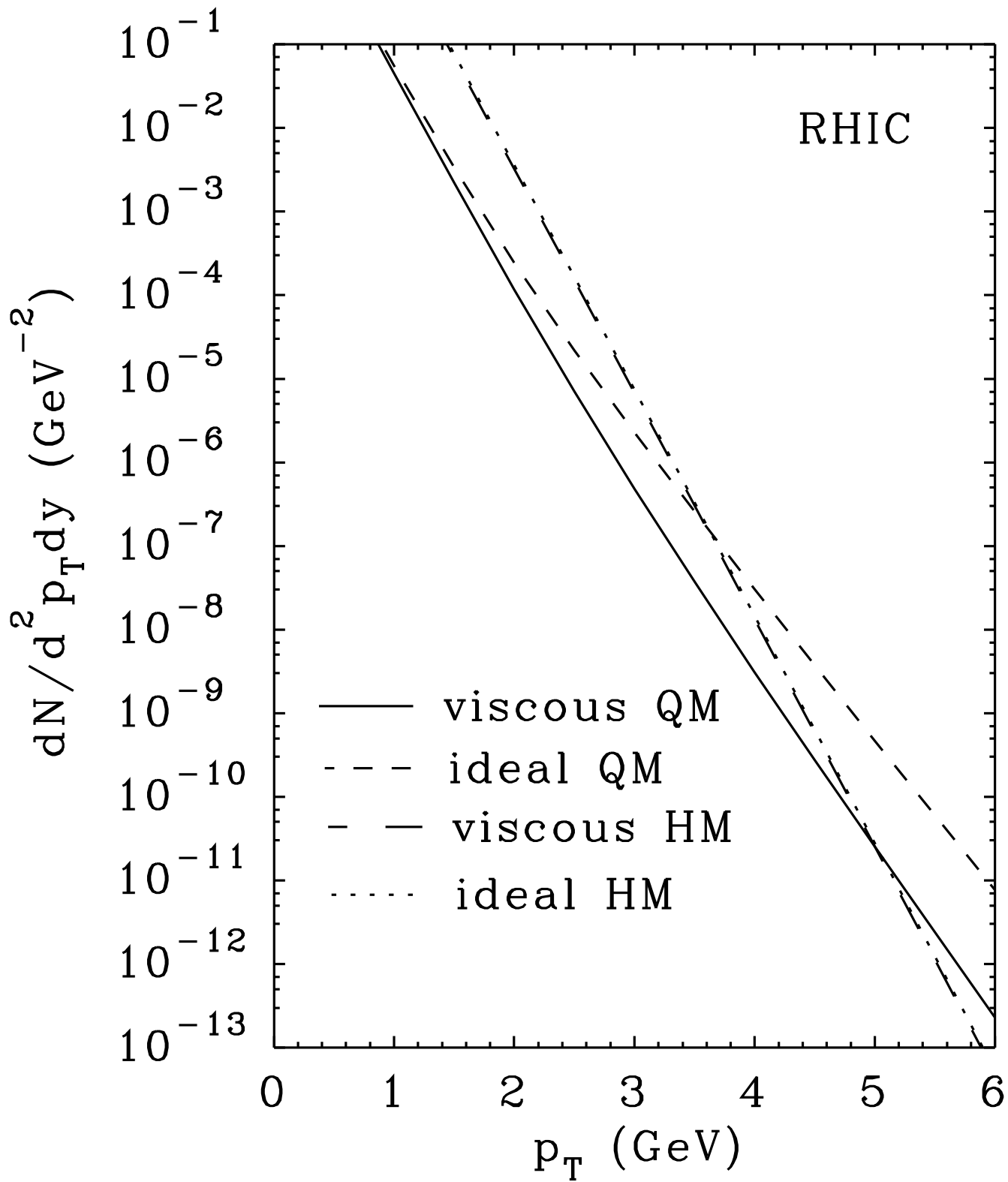
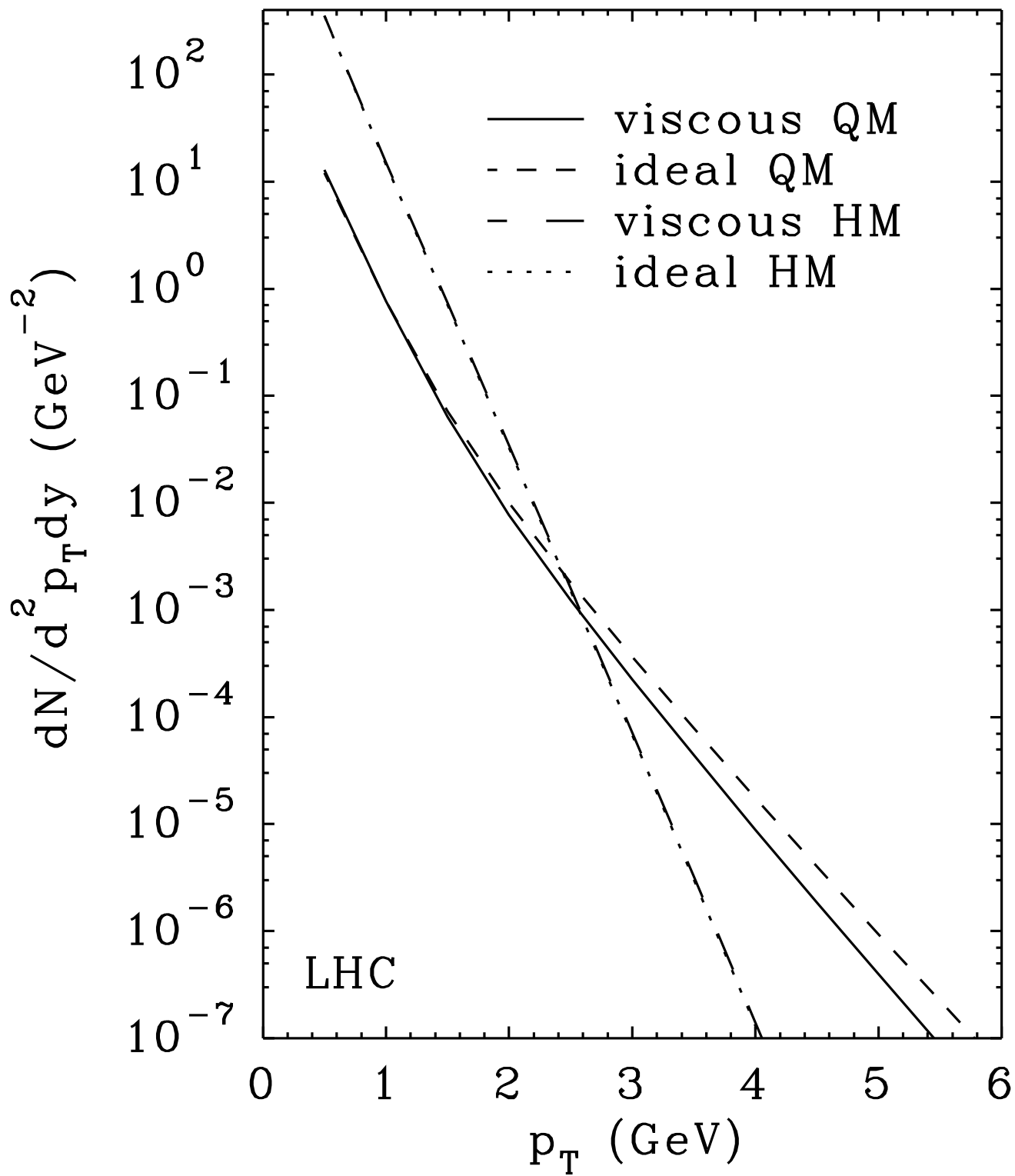
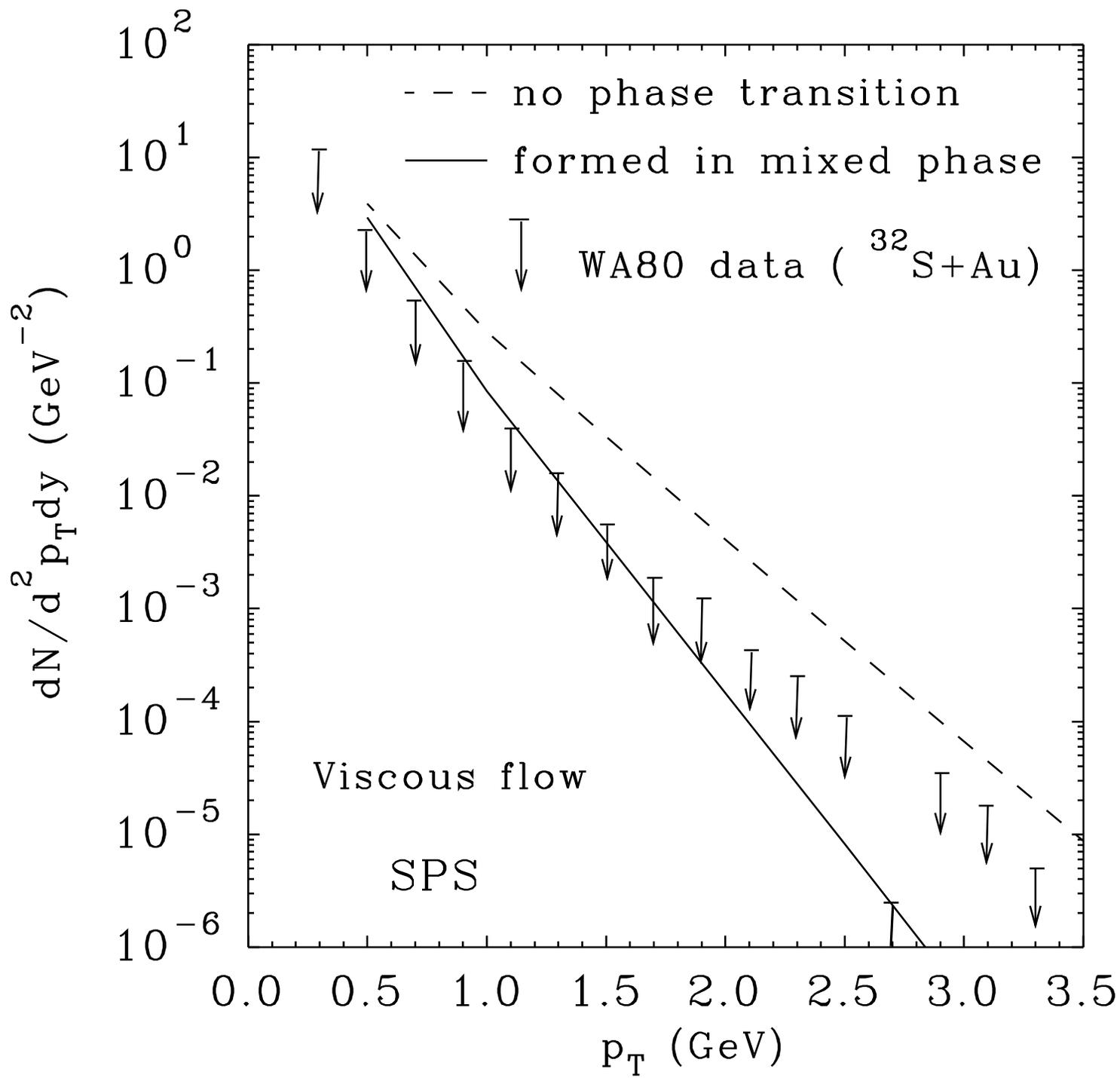
\end{document}